\documentclass[%
 reprint,
 amsmath,amssymb,
 aps,
floatfix,
]{revtex4-2}
\usepackage{array}
\usepackage{graphicx}
\usepackage{dcolumn}
\usepackage{bm}
\usepackage{multirow}

\usepackage{hyperref}
\hypersetup{
    colorlinks = true,
    linkcolor = blue,
    citecolor = blue,
    urlcolor = blue,
}
\begin{document}

\preprint{APS/123-QED}

\title{Universal Thermodynamic Topological Classes of BTZ Black Holes in Einstein and F(R) Gravity}

\author{
  Yu-Die Wan$^{1}$,
  Peng Zhao$^{1}$,
  Meng-Yao Zhang$^{2}$,
  Zheng-Wen Long$^{1}$\thanks{Corresponding author: zwlong@gzu.edu.cn}
}
\email{zwlong@gzu.edu.cn}

\affiliation{$^{1}$ College of Physics, Guizhou University, Guiyang, Guizhou 550025, People’s Republic of China}
\affiliation{$^{2}$ College of Computer and Information Engineering, Guizhou University of Commerce, Guiyang, 550014, People’s Republic of China}

\date{\today}

\begin{abstract}
In this paper, we systematically study three classes of three-dimensional Bañados-Teitelboim-Zanelli (BTZ) black holes based on different gravitational frameworks and matter field structures: the Einstein-Maxwell BTZ, the F(R)-Maxwell BTZ, and the F(R)-phantom BTZ.Within the canonical and grand canonical ensemble frameworks, we construct the generalized free energy of these black hole systems and systematically analyze the asymptotic behavior of Hawking thermodynamics.The results indicate that both gravitational modifications and matter fields significantly influence black hole topology: the F(R) gravitational modification alters the intrinsic topological class of the black hole compared to the BTZ black hole under Einsteinian gravity, while different matter fields further modulate the topological classification results. Furthermore, by comparing the two types of ensembles, we find significant differences in black hole topological classes, and these differences are correlated with the gravitational framework and matter fields. In summary, the gravitational framework, matter fields, and ensemble selection are key factors governing the topological classification of black holes; all identified topological categories belong to the existing three-dimensional spacetime topological classification system, highlighting their universality. This work deepens our understanding of the thermodynamic-topological connection in three-dimensional BTZ black holes and provides a rigorous theoretical foundation for extending such topological classifications to modified gravity theories and different matter field backgrounds.
\end{abstract}

\maketitle
\section{Introduction}
\label{sec:level1}
In recent years, the thermodynamic-topological approach has provided a novel perspective on black hole research. This method treats black holes as topological defects in thermodynamic space and defines topological invariants through the singular structure of the generalized free energy. Within this framework, the vast majority of black holes can be classified into three topological types: +1, 0, and -1,which characterize their overall stability  \cite{Wei:2022dzw}. Due to its excellent universality and computational efficiency, this method has been widely applied to various black hole scenarios, including rotating anti-de Sitter (AdS) black holes \cite{ref2,ref3}, Lovelock AdS black holes \cite{ref4,ref5}, Born-Infeld black holes \cite{ref6}, nonlinearly charged black holes \cite{ref7}, R-charged black holes \cite{ref8}, boundary matrix duals\cite{ref9}, and hairy black holes \cite{ref10}, among other complex configurations. In subsequent studies, the black hole classification system has been further refined into five topological classes and four subclasses \cite{ref11}. Ref. \cite{ref12} investigated the thermodynamic topological classification of AdS charged black holes coupled with phantom and Maxwell fields in four-dimensional dRGT massive gravity, pointing out that the statistical ensemble is the key factor determining the topological class of black holes and that there are significant differences between the results under the canonical and grand canonical ensembles. 

Among various low-dimensional black hole geometries, the three-dimensional BTZ black hole serves as an ideal system for testing gravitational theories and black hole thermodynamics \cite{ref13}. With its simple structure, exact solutions, and absence of redundant gravitational degrees of freedom, the BTZ black hole allows for a clear separation of the physical effects of modified gravity and the matter field, free from complex perturbations. Kastor, Ray, and Traschen were the first to establish that the cosmological constant \(\Lambda\) can be consistently identified as thermodynamic pressure, whose conjugate variable corresponds to the thermodynamic volume of AdS black holes \cite{ref14,ref15,ref16,ref17}. This key framework has motivated intensive studies on the thermodynamics \cite{ref18,ref19,ref20,ref21,ref22,ref23,ref24,ref25,ref27,ref28}, phase transitions \cite{ref29,ref30,ref31,ref32,ref33,ref34,ref35,ref36,ref37,ref38,ref39,ref40,ref41}, and Joule–Thomson expansions \cite{ref42,ref43,ref44,ref45,ref46,ref47,ref48,ref49,ref50} of AdS black holes within the extended phase space. Furthermore, its deep connections to string theory and the holographic principle endow its thermodynamic and topological properties with broader theoretical significance. Most importantly, the BTZ background allows for a clean and systematic comparison of how different gravitational frameworks and matter sources modulate the topological properties of black holes, providing a significant theoretical advantage for the research in this paper\cite{ref51}.

In terms of gravitational theory, F(R) gravity is one of the most physically motivated and extensively studied modification schemes in general relativity, capable of explaining the accelerated expansion of the universe without introducing additional dark components; regarding matter fields, phantom fields possess unique negative kinetic energy characteristics, inducing non-trivial spacetime structures and thermodynamic behaviors not found in conventional Maxwellian fields \cite{ref52,ref53,ref54,ref55,ref56,ref57,ref58,ref59,ref60,ref61}. The combination of these two elements simultaneously captures the dual effects of higher-order curvature corrections and exotic matter fields, making the F(R)-phantom system a powerful and physically meaningful framework for studying the thermodynamics, phase transitions, and topological classification of BTZ black holes.

Motivated by the AdS/CFT correspondence, black holes with negative cosmological constants have garnered significant attention. In various gravity theories such as Gauss-Bonnet gravity, massive gravity, quasitopological gravity, f(R, T) gravity, rainbow gravity, and others, the coupling parameters and black hole characteristics can influence the number and positions of zero points in vector spaces. However, the total topological number typically remains constant in most scenarios \cite{ref62,ref63,ref64,ref65,ref66,ref67,ref68,ref69,ref70,ref71,ref72,ref73,ref74,ref75,ref76,ref77}. Although the field of black hole thermodynamic topology is developing rapidly, systematic topological studies combining F(R) gravity, phantom fields, three-dimensional black holes, and ensemble dependence remain scarce. To fill this gap, this paper generalizes and extends existing models to systematically study three classes of three-dimensional BTZ black holes: Maxwell field black holes in Einsteinian gravity \cite{ref78}, and black holes with coupled Maxwell and phantom fields in F(R) gravity \cite{ref79}.In addition, the research results can provide a new theoretical basis for understanding the critical behavior and stability of black holes, which is of great significance for the in-depth study of black hole thermodynamics and topological field theory.

The rest of the paper is organized as follows. Section \ref{sec:level1} reviews the previous classification system of five topological classes and four subclasses, laying the theoretical foundation for the subsequent analysis. Sections \ref{sec:level3} and \ref{sec:level4} investigate Maxwell field black holes in Einsteinian gravity and black holes with coupled Maxwell and phantom fields in F(R) gravity, respectively, and discuss their topological characteristics under different statistical ensembles. Section \ref{sec:level5} summarizes the main conclusions of this paper and proposes possible extensions for future work.
\section{PRELIMINARIES: TOPOLOGICAL THERMODYNAMICS}
\label{sec:level2}
We adopt the perspective that black hole thermodynamic configurations can be interpreted as intrinsic topological defects embedded within the extended thermodynamic parameter manifold. To establish the theoretical foundation for subsequent topological classification, we first introduce the generalized off-shell Helmholtz free energy functional for a black hole thermodynamic system \cite{ref80,ref81}
\begin{equation}
    \mathcal{F} = M - \frac{S}{\tau},
    \label{eq:ziyouneng}
\end{equation}
in this context,  \(\tau\) corresponds to the inverse temperature of the cavity confining the black hole, which places the system in an off-shell state. Importantly, the shell condition is realized uniquely when \(\tau\) coincides with the inverse Hawking temperature, i.e., \(\tau\) = \(\beta\) = \(\frac{1}{T} \). In this case, the generalized off-shell free energy \(\mathcal{F}\) recovers the standard Helmholtz free energy \(F=M-TS\) \cite{ref82,ref80,ref83,ref84}. 

To systematically probe the topological defect structure of black hole thermodynamics, we construct a two-dimensional vector field defined over the parameter space spanned by the event horizon radius \(r_h\) and an auxiliary angular coordinate \(\Theta \) \cite{ref85}
\begin{equation}
    \phi = \left( \phi^{r_h}, \phi^\Theta \right) = \left( \frac{\partial \hat{\mathcal{F}}}{\partial r_h}, \frac{\partial \hat{\mathcal{F}}}{\partial \Theta} \right),
    \label{eq:fai}
\end{equation}
here \(\hat{\mathcal{F}}\) stands for the modified free energy introduced for angular boundary regularization, and the angular parameter \(\Theta \) is confined to the compact interval \([0,\pi]\). The angular component \(\phi^\Theta\) exhibits asymptotic divergence at the boundaries \(\Theta=0\) and \(\Theta=\pi\), which naturally endows the vector field with outward-pointing orientation at the parameter-space boundary and provides a self-consistent compactification scheme. The isolated zero points of the vector field \(\phi\) are exactly identified with thermodynamic critical points, phase transition thresholds, and stable/unstable equilibrium black hole solutions, serving as fundamental topological defects in the thermodynamic manifold. 

Inspired by Ref. \cite{ref86},the topological current is characterized as follows
\begin{equation}
J^{\mu}=\frac{1}{2 \pi} \epsilon^{\mu v \lambda} \epsilon_{a b} \partial_{v} n^{a} \partial_{\lambda} n^{b},
\end{equation}
here, we define \(\partial_{v}\)as the derivative with respect to \(x^{v}\), where \(x^{v}\) denotes the coordinates (t,r,\(\theta\)). The unit vector n is specified by its components \(n^{1}=\frac{ \phi^{S}}{\parallel \phi\parallel }\) and \(n^{2}=\frac{ \phi^{\theta }}{\parallel \phi\parallel }\). The conservation condition \(\partial_{\mu} J^{\mu}=0\) is required to ensure a well-defined topological current. Furthermore, the topological charge over a parameter region \(\sum\) can be expressed mathematically as
\begin{equation}
Q_{t}=\int_{\Sigma} j^{0} d^{2} x=\sum_{i=1}^{N} w_{i},
\end{equation}
the winding number \(w_i\) corresponds to the zeros of \(\phi^{a}\).The total topological charge \(Q_t\) is calculated  by summing the charges associated with individual critical points, which depend on their respective winding properties. The study of black hole thermodynamic topology has been extended to various black hole classes \cite{ref89,ref90,ref91,ref92,ref93,ref94,ref95}.

This theoretical approach provides an effective means for the systematic classification of different types of black holes. In the following, we briefly outline the five established topological classifications and their four associated subclasses, with details referred to in Refs.\cite{ref11,ref88,ref85,ref96}. 
\begin{align*}
&W^{1-},\, W^{0+},\, W^{0-},\, W^{1+},\, \bar{W}^{1+},\, \hat{W}^{1+}, \\
&\widetilde{W}^{1+},\, W^{0-\leftrightarrow 1+},\, \ddot{W}^{1-}.
\end{align*}
\begin{table*}[!htbp]
\centering
\caption{Thermodynamic properties of the black hole states for the nine topological (sub)classes of 
$W^{1-}$, $W^{0+}$, $W^{0-}$, $W^{1+}$\cite{ref88}, $W^{0-\leftrightarrow 1+}$, $\bar{W}^{1+}$, $\hat{W}^{1+}$\cite{ref85}, $\widetilde{W}^{1+}$\cite{ref96}, and $\ddot{W}^{1-}$\cite{ref11}, respectively.} 
\label{tab:1}   
\large
\renewcommand{\arraystretch}{1.3}
\resizebox{1\textwidth}{!}{%
\begin{tabular}{|c|c|c|c|c|c|c|}\hline

Topological (sub)classes & Innermost & Outermost & Low $T$ ($\beta\to\infty$) & High $T$ ($\beta\to0$) & DP & $W$ \\\hline

$W^{1-}$               & Unstable & Unstable & Unstable large & Unstable small & In pairs & $-1$ \\\hline
$W^{0+}$               & Stable   & Unstable & Unstable large + stable small & No & One more GP & $0$ \\\hline
$W^{0-}$               & Unstable & Stable   & No & Unstable small + stable large & One more AP & $0$ \\\hline
$W^{1+}$               & Stable   & Stable   & Stable small & Stable large & In pairs & $+1$ \\\hline
$W^{0-\leftrightarrow 1+}$ & Unstable & Stable & No & Stable large & One more AP & $0$ or $+1$ \\\hline
$\bar{W}^{1+}$         & Stable   & Stable   & No & Stable large & In pairs & $+1$ \\\hline
$\hat{W}^{1+}$         & Stable   & Stable   & Unstable small + two stable small & Stable large & One more GP & $+1$ \\\hline
$\widetilde{W}^{1+}$   & Unstable & Stable   & Stable small & Unstable small + stable small + stable large & One more AP & $+1$ \\\hline
$\ddot{W}^{1-}$& Unstable & Stable   & Unstable small & Unstable small + unstable small + stable large & One more AP & $-1$ \\ \hline

\end{tabular}%
}
\end{table*}
We summarize below the asymptotic behaviors of the Hawking temperatures corresponding to these distinct topological classes and subclasses, in the limiting regimes where the black hole event horizon radius \(r_h\) approaches its minimal value and tends to infinity, respectively
\begin{align}
W^{0+} &: \beta(r_m) = \infty, \quad \beta(\infty) = \infty, \label{eq:10} \\
W^{1-} ,\ddot{W}^{1-}&: \beta(r_m) = 0, \quad \beta(\infty) = \infty, \label{eq:9} \\
W^{0-}, \widetilde{W}^{1+} &: \beta(r_m) = 0, \quad \beta(\infty) = 0, \label{eq:11} \\
W^{1+}, \hat{W}^{1+} &: \beta(r_m) = \infty, \quad \beta(\infty) = 0, \label{eq:12} \\
\bar{W}^{1+},W^{0-\leftrightarrow 1+} &: \beta(r_m) = \text{fixed temperature}, \ \beta(\infty) = 0. \label{eq:13}
\end{align}

Table \ref{tab:1} summarizes the five topological categories and four subclasses of black holes, distinguishing between their innermost (smallest) and outermost (largest) states, while also characterizing their stability properties in both the low- and high-Hawking-temperature regimes.
\section{EINSTEIN–MAXWELL BTZ BLACK HOLES}
\label{sec:level3}
To investigate the thermodynamics and phase structure of three-dimensional charged black holes, we consider the minimal coupling of Einstein gravity to the Maxwell field in three-dimensional Anti-de Sitter space (AdS). Under the static, spherically symmetric ansatz, we solve the three-dimensional Einstein–Maxwell equations and obtain the exact metric of the charged BTZ black hole\cite{ref97,ref98,ref13,ref100,ref101}.
\begin{equation}
ds^2 = -f(r)dt^2 + \frac{dr^2}{f(r)} + r^2 d\varphi^2,
\end{equation}
where
\begin{equation}
f(r) = -2m + \frac{r^2}{l^2} - \frac{q^2}{2}\ln\left(\frac{r}{l}\right),
\end{equation}
where m is the mass parameter, l denotes the AdS radius, q stands for the electric charge parameter of the black hole.The corresponding thermodynamic quantities are given in Ref.\cite{ref102}.
\begin{equation}
\begin{aligned}
M &= \frac{r_h^2}{8l^2} - \frac{q^2}{16}\ln\left(\frac{r_h}{l}\right),\ 
S = \frac{1}{2}\pi r_h,\ 
T = \frac{r_h}{2\pi l^2} - \frac{q^2}{8\pi r_h},\\
\Phi &= -\frac{q}{8}\ln\left(\frac{r_h}{l}\right),\ 
Q = q,\ 
V = \pi r_h^2 - \frac{\pi}{4}q^2 l^2,\ 
P = \frac{1}{8\pi l^2},
\end{aligned}
\end{equation}
where \(r_h\) is the position of the event horizon.The following analysis delves into the universal classes of thermodynamic topology for this black hole within the framework of canonical and grand canonical ensembles, respectively.

\subsection{\label{sec:leve3A} Black holes in the context of the canonical ensemble}
In this section, we explore the universal thermodynamic topological classification of charged BTZ black holes in the Einstein-Maxwell field within the canonical ensemble framework. Based on the definition of generalized free energy in Eq.\ref{eq:ziyouneng}
\begin{equation}
  \mathcal{F} =\frac{\pi r_h}{8} \left( 2P r_h - \frac{1}{\tau} \right) - \frac{q^2}{16}\ln\bigl(2 r_h\sqrt{2\pi P}\,\bigr),  
\end{equation}
in this context,  \(\tau\) corresponds to the inverse temperature of the cavity confining the black hole, which places the system in an off-shell state. Importantly, the shell condition is realized uniquely when \(\tau\) coincides with the inverse Hawking temperature, i.e., \(\tau\) = \(\beta\) = \(\frac{1}{T} \). In this case, the generalized off-shell free energy \(\mathcal{F}\) recovers the standard Helmholtz free energy \(F=M-TS\) \cite{ref82,ref80,ref83,ref84}.

Following Eq.\ref{eq:fai}, the two-component vector field is employed
\begin{equation}
\phi^{r_h} = -\frac{q^2}{16r_h} + 2P\pi r_h - \frac{\pi}{2\tau},\phi^{\Theta} = -\cot \Theta \csc \Theta.
\end{equation}
Taking the boundary condition \(\phi^{r_h}=0\), the inverse temperature parameter \(\tau\) can be expressed as 
\begin{equation}
    \tau = \beta=\frac{8\pi r_h}{32P\pi r_h^2 - Q^2}.
\end{equation}
\begin{figure}[!htbp]
  \centering
  \includegraphics[width=0.3\textwidth]{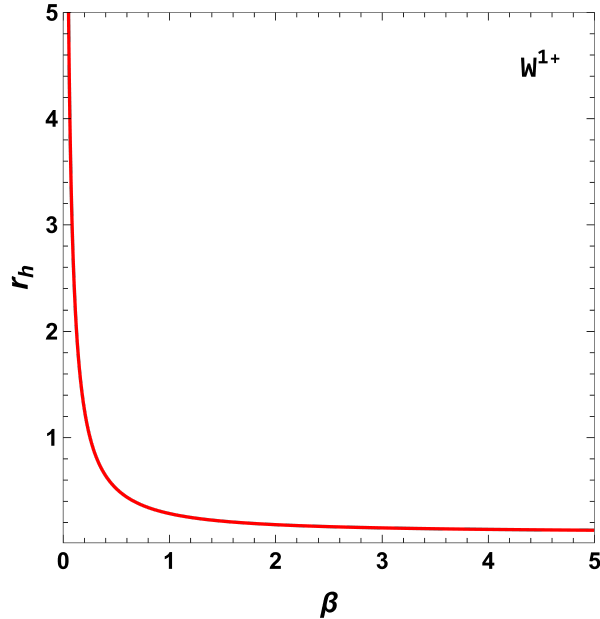}
  \caption{In the \(r_h-\beta\)  plane, the zeros of the vector \(\phi^{r_h}\) for a charged BTZ black hole (with parameters \(Q/r_0= 1\) , \(P/r_0^2= 1\) ). The stable branch is represented by the red curve}
  \label{fig:zhengzeQ=1} 
\end{figure}
\begin{figure}[!htbp]
  \centering
  \includegraphics[width=0.3\textwidth]{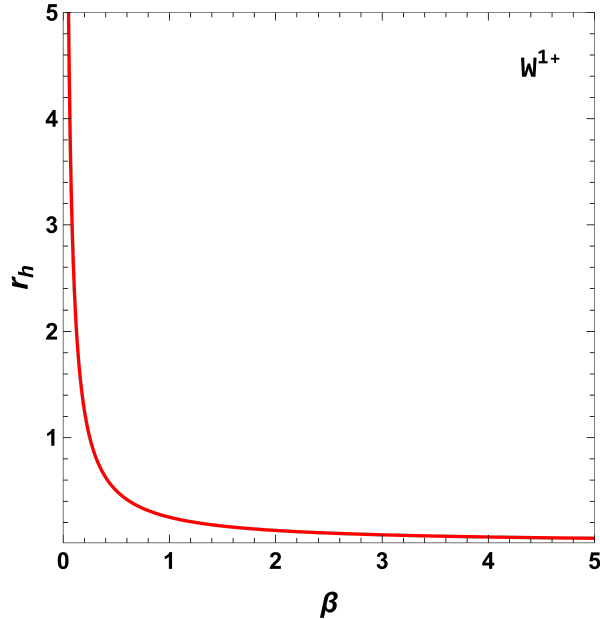}
  \caption{In the \(r_h-\beta\)  plane, the zeros of the vector \(\phi^{r_h}\) for a charged BTZ black hole (with parameters \(Q/r_0 = 0.1\), \(P/r_0^2= 1\) ). The stable branch is represented by the red curve}
  \label{fig:zhengzeQ=0.1} 
\end{figure}

Fig.\ref{fig:zhengzeQ=1} shows the zero distribution of the vector field \(\phi^{r_h}\) in the  plane for the Einstein–Maxwell charged BTZ black hole in the canonical ensemble. The curve exhibits a single continuous stable branch, with the horizon radius varying monotonically with the inverse temperature, and no unstable branch, phase transition point, or annihilation point appears across the entire temperature range. As indicated by Eq.\ref{eq:13}, the \(W^{1+}\) class satisfies the asymptotic behavior \(\beta(r_m) = \infty\) and \(\beta(\infty) = 0\), which is fully consistent with the feature of the curve in Fig.\ref{fig:zhengzeQ=1}. According to the criteria in Table \ref{tab:1} that the \(W^{1+}\) topological class is stable at both the innermost and outermost states with a topological charge \(W=+1\), this black hole is classified into the \textbf{\(W^{1+}\)} topological class. Furthermore, a comparison of Fig.\ref{fig:zhengzeQ=1} and Fig.\ref{fig:zhengzeQ=0.1} reveals that variations in the charge parameter under the canonical ensemble do not alter the topological class. Both of these conclusions are consistent with the results reported in Ref. \cite{ref78}.
\subsection{\label{sec:leve3B}Black holes in the context of the grand canonical ensemble}
In the grand canonical ensemble, we investigate the thermodynamic topological classification of the Einstein–Maxwell charged BTZ black hole. The generalized off-shell free energy and vector field are derived, 
\begin{equation}
\begin{aligned}
\mathcal{F} &= M - \frac{S}{\tau} - \Phi Q \\
& = \frac{r_h^2}{8l^2} - \frac{1}{16}q^2\ln\left(\frac{r_h}{l}\right) - \frac{\frac{\pi r_h}{2} - \frac{1}{8}qQ\tau\ln\left(\frac{r_h}{l}\right)}{\tau},
 \label{eq:15}
\end{aligned}
\end{equation}
\begin{equation}
\phi^{r_h} = \frac{Q^2}{16r_h} + 2P\pi r_h - \frac{\pi}{2\tau},
\end{equation}
\begin{equation}
    \phi^{\Theta} = -\cot \Theta \csc \Theta.
\end{equation}

And the inverse temperature \(\beta\) is obtained at the zero point of the vector field
\begin{equation}
    \tau =\beta= \frac{8\pi r_h}{Q^2 + 32 P \pi r_h^2}.
\end{equation}
\begin{figure}[!htbp]
  \centering
  \includegraphics[width=0.3\textwidth]{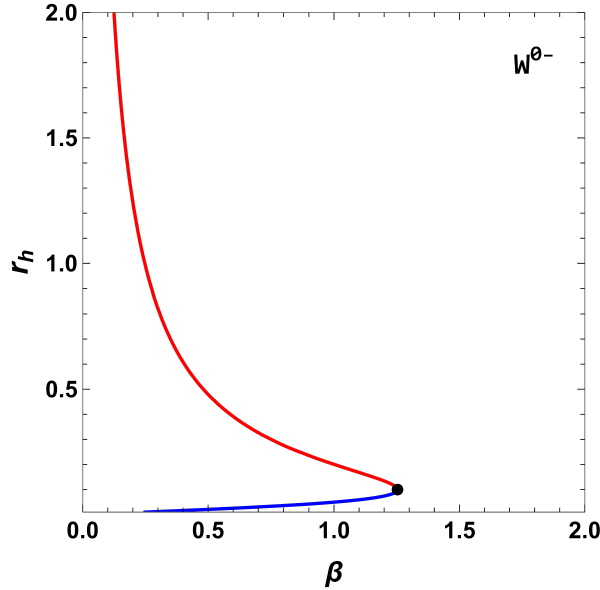}
  \caption{In the \(r_h-\beta\)  plane, the zeros of the vector \(\phi^{r_h}\) for a charged BTZ black hole (with parameters \(Q/r_0 = 1\), \(P/r_0^2=1\) ). At the intersection of the blue unstable branch (\(w = -1\)) and the red stable branch (\(w = +1\)), denoted by the annihilation point \(AP\) is marked in black}
  \label{fig:juzhengzeQ=1} 
\end{figure}
\begin{figure}[!htbp]
  \centering
  \includegraphics[width=0.3\textwidth]{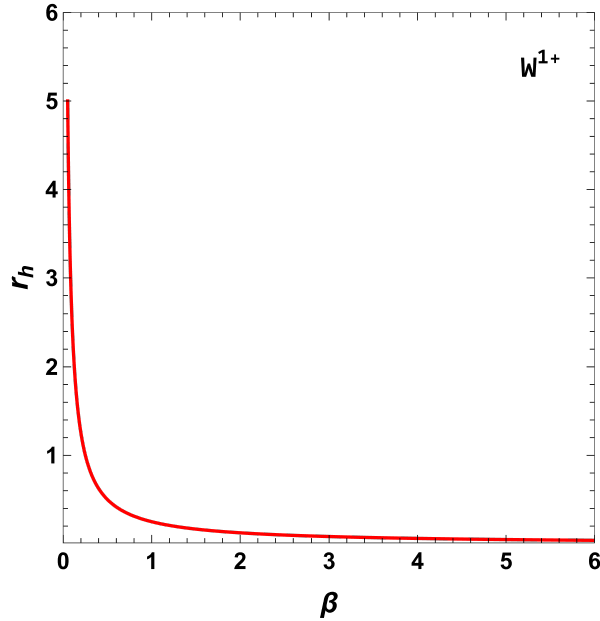}
  \caption{In the \(r_h-\beta\)  plane, the zeros of the vector \(\phi^{r_h}\) for a charged BTZ black hole (with parameters \(Q/r_0 = 0.1\), \(P/r_0^2= 1\) ). The stable branch is represented by the red curve}
  \label{fig:juzhengzeQ=0.1} 
\end{figure}
As shown in Fig.\ref{fig:juzhengzeQ=1}, the phase diagram contains a stable large-black-hole branch and an unstable small-black-hole branch, with a total topological number \(W=0\). This configuration corresponds to the topological class \(W^{0-}\), where a stable large black hole and an unstable small black hole coexist at high temperatures, and no black hole phase exists above the critical inverse temperature \(\beta_c\). When the charge parameter is small (Fig.\ref{fig:juzhengzeQ=0.1}), the horizon radius decreases monotonically with \(\beta\), and only a single stable black hole phase exists over the whole parameter range. In this case, the black hole remains stable at both minimum and maximum horizon radii, belonging to the \(W^{1+}\) topological class. 
By comparing Fig.\ref{fig:juzhengzeQ=1} and Fig.\ref{fig:juzhengzeQ=0.1}, it is clear that under the grand canonical ensemble, the charged BTZ black hole exhibits different thermodynamic topological classes for different charge parameters: the large charge corresponds to the \(W^{0-}\) class, while the small charge corresponds to the \(W^{1+}\) class. The results demonstrate that varying the charge parameter can induce a topological class transition of the black hole. 
\section{Thermodynamic Topology of F(R)-Maxwell and F(R)-Phantom BTZ Black Holes}
\label{sec:level4}
In this section, we investigate three-dimensional black holes in F(R) conformally invariant gravity coupled to a phantom field. This system belongs to conformally invariant power-Maxwell theories with a traceless energy–momentum tensor, providing an analytically tractable framework for exploring high-curvature corrections and thermodynamics of dark-energy-like matter in low-dimensional gravity. Explicit black hole solutions exist only for a constant negative scalar curvature \(R_0<0\), which corresponds to asymptotically AdS spacetime provided \(R_0=6\Lambda \). 
The static circularly symmetric metric takes the standard BTZ form 
\begin{equation}
    ds^2 = -\psi(r) dt^2 + \frac{dr^2}{\psi(r)} + r^2 d\varphi^2,
\end{equation}
with the analytic metric function
\begin{equation}
    \psi(r) = -m_0 - \frac{R_0 r_h^2}{6} - \frac{\eta q}{2^{1/4} \left(1 + f_{R_0}\right) r_h},
\end{equation}
here, \(m_0\) is the mass parameter, \(r_h\) is the event horizon radius, \(q\) the electric charge, \(\eta=\pm1\) distinguishes the Maxwell (\(\eta=+1\)) and phantom (\(\eta=-1\)) fields, and \( f_{R_{0}}=\left.(d f / d R)\right|_{R=R_{0}}\) is the F(R) gravity coupling. The spacetime is regular everywhere except for a curvature singularity at r=0 and is asymptotically AdS. Using the horizon condition \(\psi(r_h)=0\), we directly derive analytic expressions for the total mass, entropy, electromagnetic potential, and electric charge
\begin{equation}
\begin{aligned}
M &= -\frac{(1+f_{R_0})R_0 r_h^2}{48} - \frac{\eta q}{2^{13/4} r_h},
&
S &= \frac{\pi (1+f_{R_0}) r_h}{2}, \\[4pt]
U &= -\frac{q^{2/3}}{r_h},
&
Q &= \frac{3 q^{1/3}}{2^{13/4}}.
 \label{eq:26}
\end{aligned}
\end{equation}
The \(F(R)\)-phantom \(BTZ\) black hole possesses well-defined thermodynamic potentials, a stable horizon structure, and distinct stability regimes, providing an ideal arena for thermodynamic topological classification. 
\subsection{\label{sec:leve4A}Black holes in the context of the canonical ensemble}
In this section, we investigate AdS black holes coupled with Maxwell and phantom fields in F(R) gravity, and analyze their thermodynamic topological properties within the canonical ensemble. The corresponding generalized free energy can be expressed as follows: 
\begin{equation}
\mathcal{F} = - (1 + f_{R_0}) \left( \frac{R r_h^2}{48} + \frac{\pi r_h}{2\tau} \right) - \frac{q \eta}{2^{13/4} r_h},
\end{equation}

\begin{figure}[!htbp]
  \centering
  \includegraphics[width=0.3\textwidth]{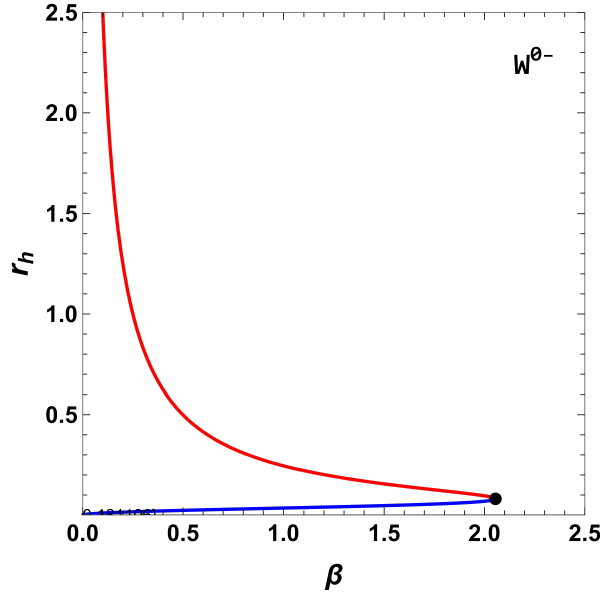}
  \caption{In the \(r_h-\beta\) plane, the zeros of the vector \(\phi^{r_h}\) for a charged BTZ black hole for Maxwell (\(\eta=+1\)) with parameters \(f_{R_0}=1\) ,\(Q/r_0 = 0.1\) ,\(P/r_0^2= 1\). At the intersection of the blue unstable branch (\(w=-1\)),and the red stable branch (\(w=+1\)), denoted by the annihilation point (\(AP\)) is marked in black}
  \label{fig:F(R)zhengzemaikesiweiQ=0.1} 
\end{figure}
\begin{figure}[!htbp]
  \centering
  \includegraphics[width=0.3\textwidth]{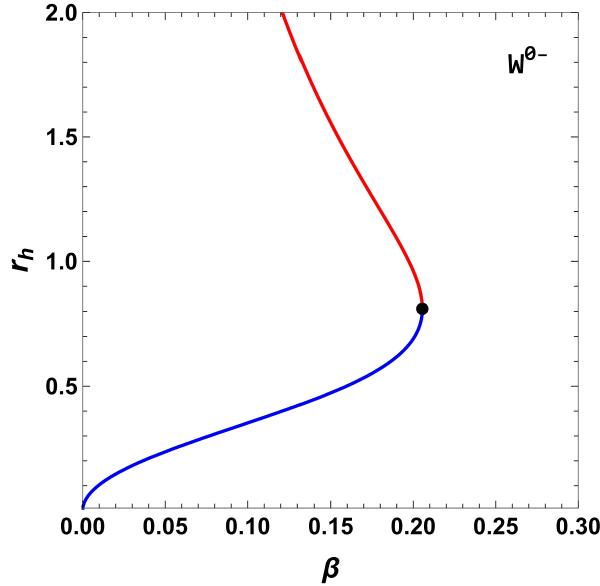}
  \caption{In the \(r_h-\beta\)  plane, the zeros of the vector \(\phi^{r_h}\) for a charged BTZ black hole for Maxwell (\(\eta=+1\))(with parameters (\(f_{R_0} = 1\) ,\(Q/r_0 = 1\),\(P/r_0^2= 1\) ). At the intersection of the blue unstable branch (\(w = -1\)) and the red stable branch (\(w = +1\)), denoted by the annihilation point (AP) is marked in black}
  \label{fig:F(R)zhengzemaikesiweiQ=1}
\end{figure}
\begin{figure}[!htbp]
  \centering
  \includegraphics[width=0.3\textwidth]{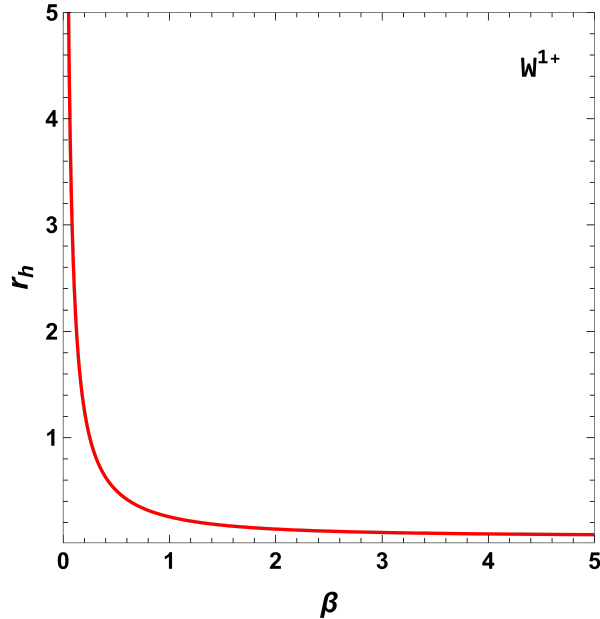}
  \caption{In the \(r_h-\beta\)  plane, the zeros of the vector \(\phi^{r_h}\) for a charged BTZ black hole for phantom (\(\eta=-1\))(with parameters (\(f_{R_0} = 1\) ,\(Q/r_0 = 0.1\),\(P/r_0^2= 1\) ). The stable branch is represented by the red curve}
  \label{fig:F(R)zhengzeyoulingQ=0.1}
\end{figure}
\begin{figure}[!htbp]
  \centering
  \includegraphics[width=0.3\textwidth]{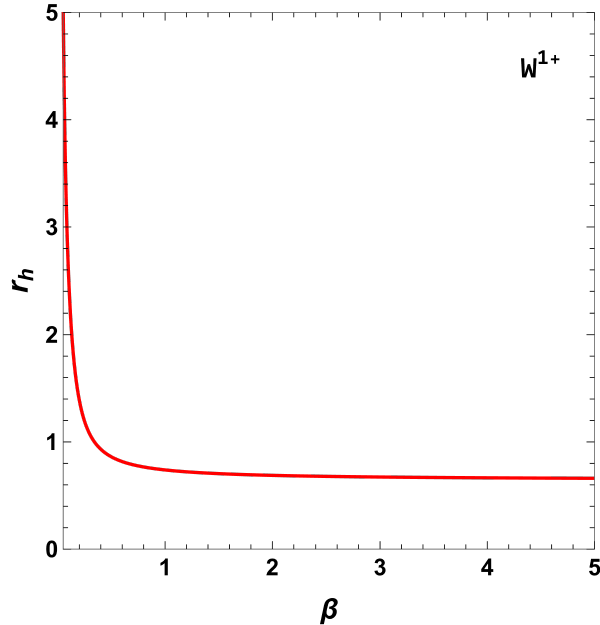}
  \caption{In the \(r_h-\beta\)  plane, the zeros of the vector \(\phi^{r_h}\) for a charged BTZ black hole for phantom (\(\eta=-1\))(with parameters (\(f_{R_0}= 1\) ,\(Q/r_0 = 1\),\(P/r_0^2= 1\) ). The stable branch is represented by the red curve}
  \label{fig:F(R)zhengzeyoulingQ=1}
\end{figure}
thus, the vector components can be written as
\begin{equation}
\begin{aligned}
 \phi^{r_h}  &=- \frac{(1 + f_{R_0}) R r_h}{24} + \frac{q \eta}{2^{13/4} r_h^2} - \frac{(1 + f_{R_0}) \pi}{2\tau},\\
 \phi^{\theta} &=-\cot \Theta \csc \Theta.
\end{aligned}
\end{equation}
and
\begin{equation}
    \tau=\beta=\frac{27 (1 + f_{R_0}) \pi r_h^2}{108 (1 +f_{R_0}) \pi P r_h^3 + 128 \sqrt{2} Q^3 \eta},
\end{equation}

Figs.\ref{fig:F(R)zhengzemaikesiweiQ=0.1} and \ref{fig:F(R)zhengzemaikesiweiQ=1} clearly show that, for the Maxwell field (\(\eta=+1\)), the black hole remains in the topological class \(W^{0-}\) irrespective of variations in the charge parameter, with no topological phase transition occurring. Similarly, Figs.\ref{fig:F(R)zhengzeyoulingQ=0.1} and \ref{fig:F(R)zhengzeyoulingQ=1} demonstrate that, for the phantom field (\(\eta=-1\)), modifying the charge parameter also leaves the topological classification unchanged, with all solutions belonging to the class \(W^{1+}\). A combined comparison of all four figures leads to the definitive conclusion: within the canonical ensemble, the type of matter field constitutes the key factor driving the topological class transition of black holes—switching from the Maxwell field to the phantom field changes the topological class from \(W^{0-}\) to \(W^{1+}\). 
\subsection{\label{sec:leve4B}Black holes in the context of the grand canonical ensemble}
In this section, we extend our analysis of \(AdS\) black holes coupled with Maxwell and phantom fields in \(F(R)\) gravity, presented in the previous section, to the grand canonical ensemble. The corresponding generalized free energy reads
\begin{equation}
\mathcal{F} = - \frac{(1 + f_{R_0}) R r_h^2}{48} - \frac{(1 + f_{R_0}) \pi r_h}{2\tau} + \frac{q(3 - \eta)}{2^{13/4} r_h},
 \label{eq:27}
\end{equation}
\begin{figure}[!htbp]
  \centering
  \includegraphics[width=0.3\textwidth]{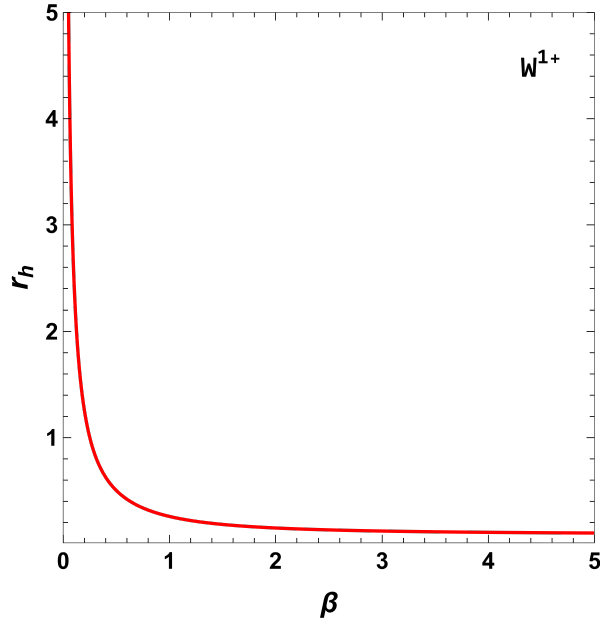}
  \caption{In the \(r_h-\beta\)  plane, the zeros of the vector \(\phi^{r_h}\) for a charged BTZ black hole for Maxwell (\(\eta=+1\))(with parameters (\(f_{R_0} = 1\) ,\(Q/r_0 = 0.1\),\(P/r_0^2= 1\) ). The stable branch is represented by the red curve}
  \label{fig:F(R)juzhengzemaikesiweiQ=0.1}
\end{figure}
\begin{figure}[!htbp]
  \centering
  \includegraphics[width=0.3\textwidth]{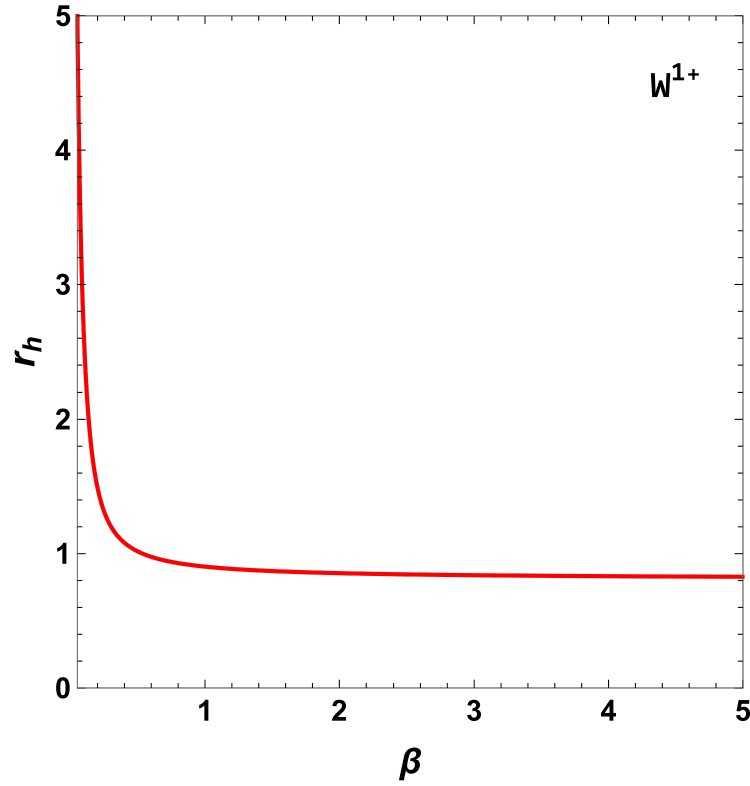}
  \caption{In the \(r_h-\beta\)  plane, the zeros of the vector \(\phi^{r_h}\) for a charged BTZ black hole for Maxwell (\(\eta=+1\))(with parameters (\(f_{R_0} = 1\) ,\(Q/r_0 = 1\),\(P/r_0^2= 1\) ). The stable branch is represented by the red curve}
  \label{fig:F(R)juzhengzemaikesiweiQ=1}
\end{figure}

\begin{figure}[!htbp]
  \centering
  \includegraphics[width=0.3\textwidth]{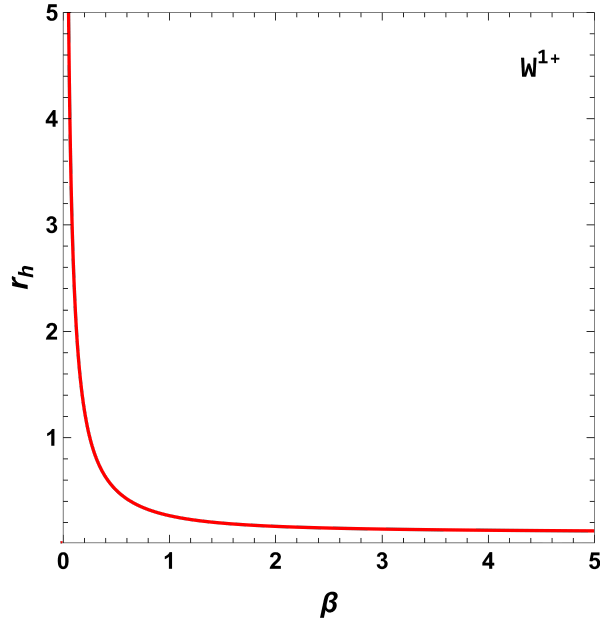}
  \caption{In the \(r_h-\beta\)  plane, the zeros of the vector \(\phi^{r_h}\) for a charged BTZ black hole for phantom (\(\eta=-1\))(with parameters (\(f_{R_0} = 1\) ,\(Q/r_0 = 0.1\),\(P/r_0^2= 1\) ). The stable branch is represented by the red curve}
  \label{fig:F(R)juzhengzeyoulingQ=0.1}
\end{figure}
\begin{figure}[!htbp]
  \centering
  \includegraphics[width=0.3\textwidth]{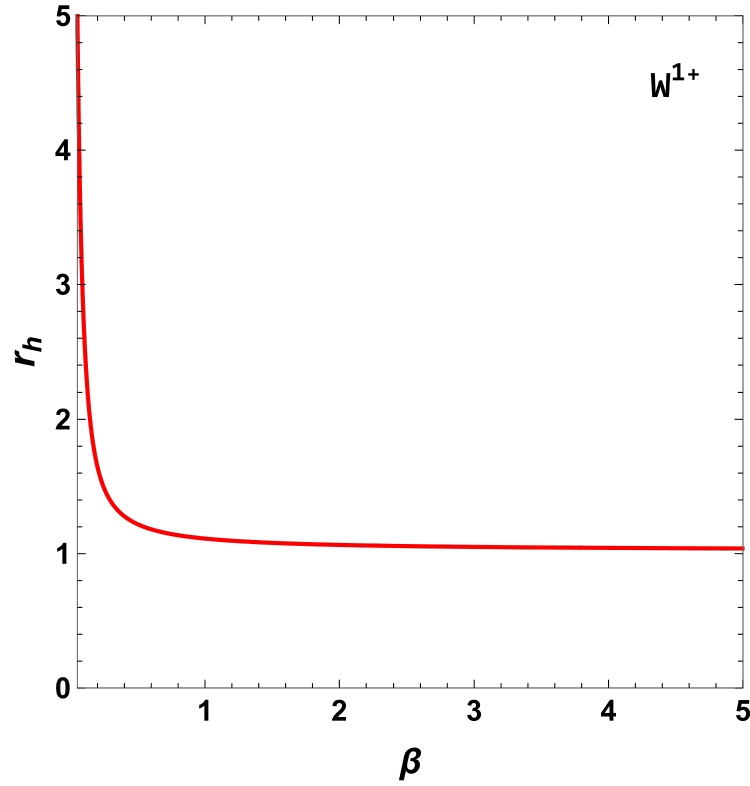}
  \caption{In the \(r_h-\beta\)  plane, the zeros of the vector \(\phi^{r_h}\) for a charged BTZ black hole for phantom (\(\eta=-1\))(with parameters (\(f_{R_0} = 1\) ,\(Q/r_0 = 1\),\(P/r_0^2= 1\) ). The stable branch is represented by the red curve}
  \label{fig:F(R)juzhengzeyoulingQ=1}
\end{figure}
the following equation provides the zero points of the component \(\phi\)
\begin{equation}
\begin{aligned}
\phi^{r_h}& =\frac{q(\eta - 3)}{2^{13/4} r_h^2} - \frac{(1 + f_{R_0})\pi}{2\tau} - \frac{(1 + f_{R_0}) r_h R}{24},\\
\phi^{\theta}&=-\cot \Theta \csc \Theta.
\end{aligned}
\end{equation}
Thus
\begin{equation}
    \tau=\frac{27 (1 + f_{R_0}) \pi r_h^2}{4\left(27 (1 + f_{R_0}) \pi P r_h^3 + 32 \sqrt{2} Q^3 (\eta - 3)\right)}.
\end{equation}
Analogous to the analysis in Section \ref{sec:leve3B}. From the analysis of the \(r-\beta\) diagrams in fig.\ref{fig:F(R)juzhengzemaikesiweiQ=0.1} and fig.\ref{fig:F(R)juzhengzemaikesiweiQ=1}, we find that varying only the charge parameter does not alter the topological class of black holes, which all belong to the \(W^{1+}\) class.
\begin{table*}[!htbp]
\centering
\caption{The universal thermodynamic topological classifications of the Charged BTZ charged black holes and F(R) BTZ black holes.}
\label{tab:2}
\renewcommand{\arraystretch}{1.2} 
\begin{tabular}{|c|c|c|c|c|}
\hline
BH solutions & Ensemble & Material field & Charge & W classes \\
\hline
\multirow{4}{*}{Charged BTZ BH} 
& \multirow{2}{*}{grand canonical} & $\eta=1$ & $Q=0.1$ & $W^{1+}$ \\
& & & $Q=1$ & $W^{0-}$ \\
\cline{2-5}
& \multirow{2}{*}{canonical} & $\eta=1$ & $Q=0.1$ & $W^{1+}$ \\
& & & $Q=1$ & $W^{1+}$ \\
\hline
\multirow{6}{*}{F(R)-Charged BTZ BH}
& \multirow{2}{*}{grand canonical} & $\eta=1$ & $Q=0.1$ & $W^{1+}$ \\
& & & $Q=1$ & $W^{1+}$ \\
\cline{2-5}
& & $\eta=-1$ & $Q=0.1$ & $W^{1+}$ \\
& & & $Q=1$ & $W^{1+}$ \\
\cline{2-5}
& \multirow{2}{*}{canonical} & $\eta=1$ & $Q=0.1$ & $W^{0-}$ \\
& & & $Q=1$ & $W^{0-}$ \\
\cline{2-5}
& & $\eta=-1$ & $Q=0.1$ & $W^{1+}$ \\
& & & $Q=1$ & $W^{1+}$ \\
\hline
\end{tabular}
\end{table*}
Similarly, the investigation of Fig.\ref{fig:F(R)juzhengzeyoulingQ=0.1} and Fig.\ref{fig:F(R)juzhengzeyoulingQ=1} shows that changing only the matter field also leaves the topological classification of black holes unchanged, with all solutions still categorized into the \(W^{1+}\) class.
\section{Conclusions}
\label{sec:level5}
In this work, we systematically investigate the thermodynamic topological classification of three-dimensional BTZ black holes coupled with conformally invariant Maxwell and phantom fields in F(R) gravity, within both the canonical and grand canonical ensembles. We also performed a direct comparison with the topological behavior of charged AdS black holes in Einstein gravity, revealing rich and distinct topological features induced by modified gravity, matter fields, and the choice of ensemble. All relevant results are systematically summarized in Table \ref{tab:2}.

For the Einstein–Maxwell BTZ black hole, varying the charge parameter does not alter the topological class in the canonical ensemble, which is fully consistent with the existing results. In the grand canonical ensemble, by contrast, a change in charge can modify the topological class. For the F(R)-Maxwell and F(R)-phantom BTZ black holes , the topological class in the canonical ensemble is determined by the type of matter field: the Maxwell field corresponds to \(W^{0-}\), while the phantom field corresponds to \(W^{1+}\), and the charge has no effect on the classification. In the grand canonical ensemble, neither the charge nor matter field changes the topological class, and all solutions uniformly fall into the \(W^{1+}\) class.

Further comparison between the two black hole models shows that the gravitational background significantly affects the topological classification. In the canonical ensemble, the difference between F(R) gravity and Einstein gravity directly leads to distinct topological classes of black holes. In the grand canonical ensemble, a change in gravitational background also triggers a topological transition at \(Q=1\), but has no obvious effect at \(Q=0.1\).

The gravitational background considered in this work is Einstein gravity and F(R) gravity. Our results indicate that modifications of gravitational theories can significantly regulate the thermodynamic topological classes of black holes. On this basis, future work can be extended to more general modified gravity frameworks, such as higher-dimensional gravity, Horndeski gravity, and scalar–tensor gravity, to systematically explore the influence of different gravitational structures on black hole topological classes. In addition, all topological classes obtained in this work do not go beyond the existing universal classification system, which directly verifies the strong universality of the black hole thermodynamic topological classification framework and provides reliable support for extending this classification method to a wider range of gravitational and matter field models. 
\begin{acknowledgments}
This work was supported by the National Natural Science Foundation of China (No.12265007), and the Guizhou Provincial Major Scientific and Technological Program (XKBF(2025)010).
\end{acknowledgments}
\nocite{*}
\bibliographystyle{unsrt}
\bibliography{newtemp}

\end{document}